\documentclass[twocolumn, showpacs]{revtex4}

\usepackage{graphicx, color}
\usepackage[normalem]{ulem}
\usepackage{verbatim}

\begin{document}

\title{Coercivity reduction in a two-dimensional array of nano-particles}

\author{M. Morales-Meza$^1$, P.P. Horley$^1$, A. Sukhov$^{1, 2}$, J. Berakdar$^2$}

\address{$^1$Centro de Investigaci\'{o}n en Materiales Avanzados, S.C. (CIMAV), Chihuahua/Monterrey, 31109 Chihuahua, Mexico\\
$^2$Institut f\"ur Physik, Martin-Luther-Universit\"at Halle-Wittenberg, 06120 Halle (Saale), Germany}

\date{\today}

\begin{abstract}
We report on theoretical investigation of the magnetization reversal in two-dimensional arrays of ferromagnetic nano-particles with parameters of cobalt. The system was optimized for achieving the lowest coercivity in an array of particles located in the nodes of triangular, hexagonal and square grids. Based on the numerical solution of the non-stochastic Landau-Lifshitz-Gilbert equation we show that each particle distribution type is characterized with a proper optimal distance, allowing to lower the coercivity values for approximately 30\% compared with the reference value obtained for a single nano-particle. It was shown that the reduction of coercivity occurs even if the particle position in the array is not very precise. In particular, the triangular particle arrangement maintained the same optimal distance between the particles under up to 20\% random displacements of their position within the array.
\end{abstract}


\maketitle

\section{Introduction}

Research of nano-scale magnetism was greatly catalyzed with the discovery of the Giant Magnetoresistance \cite{GrSc86,Baibich1988} leading to unprecedented progress in information storage technology \cite{Kaitsu2006}. The impressive perspectives in enhancing non-volatile magnetic memory modules \cite{Sousa2005}, micro-wave generators of GHz frequency \cite{Kiselev2003, Kaka2005} and nano-scale magnetic sensors led to the development of the new promising field -- spintronics \cite{Zutic2004}. The simplest spintronic devices, spin valves, were studied in detail (see, e.g., \cite{Kiselev2003,  Stiles2006, Silva2008}). These devices consist of a thick polarizer and a thin analyzer layers, separated by a non-magnetic spacer. The magnetization of the analyzer layer can be efficiently controlled by the external magnetic field or injected spin-polarized current, allowing to achieve magnetization reversal \cite{Fricke2012}, steady magnetization precession \cite{Silva2008} as well as magnetization relaxation to so-called canted states \cite{Bazaliy2004}.

Modeling of the magnetization dynamics usually relies on the assumption that nano-sized magnetic particles can be characterized by a single magnetic moment - a macrospin \cite{Xiao2005}. Larger objects should be treated as a many-body problem within the framework of micromagnetic simulations \cite{Scholz2003}. The latter approach allows studying domain dynamics as well as nucleation, propagation and annihilation of vortices \cite{HuSc98,Perry2007, Dantas2011}. However, it is much computationally intensive that complicates the study of large arrays consisting of thousands of particles. Due to this, many publications focus on simulations of moderate-size arrays representing the existing systems such as granular media for magnetic recording \cite{Braun2009, Berger2012}. One of the issues that greatly influences the magnetization dynamics is the long-range dipole-dipole interaction between the particles, which, in turn, strongly depends on the geometry of the system and distances between the nano-particles.

Therefore, we consider it timely and important to perform a thorough comparative study of ferromagnetic arrays with different number of particles arranged into different types of grids with the aim to optimize their number and inter-particle distance to achieve new characteristics promising for spintronic device applications. The particle arrays can be created using available lithographic processes \cite{Cao04}. We assume that the particles are not overlapping so that the interaction between them includes only the dipole-dipole term. A particular attention was paid to the robustness of the system regarding possible random displacements of the particles that can occur due to imperfect control of their growth conditions.

\section{Theoretical model}
\subsection{Single nano-particle}
Each ferromagnetic nano-particle is assumed to have a cylindric shape with a circular base with diameter of $a=6$~nm located in the $xy$-plane. The height of the cylinder $h=2.2$~nm is considered to be parallel to the $z$-axis. The volume of such nano-particle scales then as $\pi a^2 h$.

Due to the non-spherical form the present nano-particle will have non-negligible demagnetizing-field contributions\cite{Osbo45,Brow62}, the calculations of which might be a non-trivial task in general \cite{Coey10}. The demagnetizing factors can be obtained \cite{BeGr03,BeGr05} for a general cylinder with an elliptic base \cite{BeGr05a}. To simplify the calculations, we can use the formula for the demagnetizing factor along the $z$-axis of an oblate ellipsoid (which has a difference below $5\%$ relative to the exact solution given in Fig. 3 of Ref. \cite{BeGr06} for a thickness below $5$~[nm]), giving us the formula
\begin{equation}
\displaystyle N_{\mathrm{z}} = \frac{k^2}{k^2-1}\left[1-\frac{\arcsin\left(\frac{\sqrt{k^2-1}}{k}\right)}{\sqrt{k^2-1}}\right],
\label{eq_0_a}
\end{equation}
with the parameter $k=a/h$. For the values of $a$ and $h$ and due to the axial symmetry we find $N_{\mathrm{z}}\approx 0.61$, $N_{\mathrm{x}}=N_{\mathrm{y}}\approx 0.19$. This allows us in the first approximation to neglect the demagnetizing factors in $xy$-plane and to model the considered nano-particle as an infinitely large plane.

Another important issue concerns the limit size of a nano-particle to which the magnetization motion is uniform and no magnetic vortices are formed \cite{HoHe03,ChHe05}. For this we should minimize the sum of exchange- and magnetostatic energies for the core and the uniformly magnetized cylinder \cite{BeGr05} (sec. 5). The resulting expression yields a critical radius below which the magnetization rotation is uniform. From the solution of the transcendental equation
\begin{equation}
\displaystyle x = \frac{2}{1-N_{\mathrm{z}}} + \frac{2N_{\mathrm{z}}}{1-N_{\mathrm{z}}}\ln x
\label{eq_0_b}
\end{equation}
we obtain for the critical radius $\displaystyle a_{\mathrm{crit}}=\sqrt{x_{\mathrm{crit}}\frac{2A}{\mu_0 M^2_{\mathrm{S}}N_{\mathrm{z}}}}\approx 3$~nm, where $A=31\cdot 10^{-12}$~ J/m \cite{Coey10} is the exchange stiffness for bulk cobalt and $x_{\mathrm{crit}}\approx 0.21$ is the solution of eq. (\ref{eq_0_b}).

It is necessary to note that we have chosen the particle size of the same order of magnitude with the critical value, so that the macrospin approximation is still valid. At the same time the particles are reasonably large to diminish the influence of the particle's surface on its magnetization dynamics.

\subsection{Arrays of nanoparticles}

Let us consider an array of ferromagnetic particles, each characterized by a magnetization vector $\vec{M_i}$. In the framework of the macrospin approximation \cite{Xiao2005}, the magnitude of $\vec{M}_i$ for every particle is constant and equal to the saturation magnetization $M_S$, which makes it convenient to work with the normalized magnetization $\vec{m}_i = \vec{M}_i / M_S$. The magnetization dynamics of a macrospin obeys the Landau-Lifshitz-Gilbert (LLG) equation \cite{LaLi35,Gilb55}
\begin{equation}
\frac{d\vec{m}_i}{dt} = - \frac{\gamma \mu_0}{1+\alpha^2} \vec{m}_i\times \left[\vec{H}_{i EFF}(t) + \alpha \left[\vec{m}_i \times \vec{H}_{i EFF}(t)\right]\right],
\label{eq_1}
\end{equation}
where $\gamma=1.76\cdot 10^{11}$~1/(Ts) denotes the gyromagnetic ratio, $\mu_0=4\pi\cdot 10^{-7}$~ Vs/(Am) stands for the magnetic permeability and $\vec{H}_{i EFF}(t)$ is the total effective field. A ferromagnetic body with such a geometry will be characterized with the anisotropy field $\vec{H}_{i ANI} = \{H_K m_{i \mathrm{x}}, 0,0\}$ ($H_K = 2 K_1 /(\mu_0 M_{\mathrm{S}}$), as well as a demagnetizing factor of a thin film $\vec{H}_{i DEM} = \{0,0,-M_S m_{i \mathrm{z}}\}$. For the case of non-overlapping particles the Maxwell's equations suggest the dipole-dipole field
\begin{equation}
\vec{H}_{i DDI} = -\frac{V_i M_{\mathrm{S}}}{4 \pi}\sum_{j \neq i}\left[ \frac{\vec{m}_i}{r_{ij}^3} - 3\frac{(\vec{m}_i \cdot \vec{r}_{ij})\vec{r}_{ij}}{r_{ij}^5} \right],
\label{eq_2}
\end{equation}
where the subscripts $j \neq i$ denote the interacting macrospins located at the distance $r_{ij}$ from each other. As we consider a system composed of the uniform particles, their volume will be constant $V\equiv V_i$. Taking into account all the aforementioned contributions, the total effective field acting on the $i$-th ferromagnetic particle can be written as
\begin{equation}
\vec{H}_{i EFF}(t) = \vec{H}_{EXT} + \vec{H}_{i ANI} + \vec{H}_{i DEM} + \vec{H}_{i DDI},
\label{eq_3}
\end{equation}
with an external field $\vec{H}_{EXT}$ used to trigger the magnetization reversal.

To investigate the influence of the particle distribution on the properties of the system, we studied two-dimensional arrays with triangular, hexagonal and square particle arrangements (Fig. 1) characterized by the grid parameter $d$. For every geometry it was assumed that the easy axes of the particles related to the magnetocrystolline anisotropy are aligned parallel to the $x$-axis.
\begin{figure*}
\includegraphics[scale=0.85]{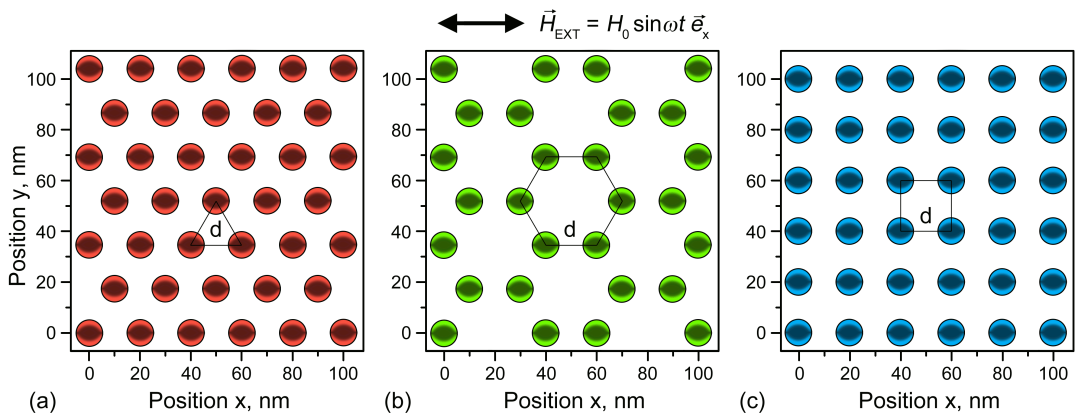}
\label{Fig1}
\caption{(Color online) Particle distributions considered in the paper (top view): a) triangular, b) hexagonal and c) square. The inter-particle distance $d$ and the unitary cell is marked for each grid type.}
\end{figure*}

The type of particle arrangement sets constrains on their magnetization dynamics, primarily due to distinct coordination numbers. For example, for a triangular grid every particle located inside the array has six nearest neighbors. The square grid has four nearest neighbors, whereas a particle located in a vertex of a hexagon has only three neighbors (Fig. 1). Due to this, one can expect that triangular grid will be more magnetically stiff in comparison with hexagonal and square particle arrangements. The particles located at the perimeter of the array have an incomplete set of the neighbors, which leads to less constrained dynamics. The ratio of ``perimeter" to ``interior" particle number can be varied by changing the size of the array. To simplify the analysis we introduced a single array size parameter $N$, assuming that the particles are located in the nodes of $N \times N$ grid. This is a straightforward approach for square and triangular grids (for the latter, each second line is shifted), so that the particle number in both cases is $N_P = N^2$. Hexagonal grid can be obtained from the triangular one by omission of certain sites, so that the total number of particles in this case is $N_P = \frac{2}{3}N^2$. For the cases of square and triangular particle arrangement, the percentage of perimeter particles is defined as $4(N-1)/N^2$. The ranges of $N$ were chosen in the way allowing to vary the aforementioned ratio from $64\%$ ($N = 5$) to $8.7\%$ ($N = 45$), so that we studied the marginal cases when the number of perimeter sites is either dominant (over 50$\%$) or almost negligible (under 10$\%$).

For the large particle arrays, the number of perimeter sites is small so that one may expect considerable uniformity in magnetization dynamics. Since the total magnetization of the system is calculated as an average over all macrospins $\vec{m} = \Sigma_i \vec{m}_i/N_P$, it will produce smoother hysteresis curves for a large $N_P$ by assigning lower weight to fluctuations of individual macrospins. However, an array containing thousands of particles will have a considerable size -- for example, a grid formed by $50 \times 50$ particles spaced at 20 nm yields the total area of $1 \times 1 \mu$m$^2$, which is beyond the nano-scale range. In the opposite case of a very small particle array the dominant contribution of perimeter sites can possibly degrade the overall system performance.

\section{Numerical results and discussion}

For our calculations we assumed nano-particles of cobalt with parameters reported from the spin valve experiments \cite{Kiselev2003}: $H_K = 500$ Oe, $M_S = 18.2$ kOe and $\alpha = 0.014$. The Eq. (3) at zero Kelvin was solved with the Runge-Kutta method of the 4th degree \cite{Press2007} with a time step $\tau_0 = 0.05$ ps required to achieve the sufficient accuracy of magnetization dynamics. The macrospin reversal was triggered by harmonically-varying magnetic field $\vec{H}_{EXT} = H_0 \sin(\omega t) \vec{e}_{\mathrm{x}}$ applied along the $x$-axis with the frequency $\omega/(2\pi) = 0.5$~GHz and the amplitude $H_0 = 2.2 \times 10^5$~A/m. The $m_{\mathrm{x}}$ component of the averaged total magnetization features a clear hysteresis saturating at $m_{\mathrm{x}} = m/M_{\mathrm{s}} \to \pm 1$ (Fig. 2).

The main parameters defining the magnetization reversal are the remanence $m_{x0}$ and the coercivity $H_C$, calculated at $H_{EXT} = 0$ and $m_x = 0$, respectively. The system with higher remanence will have better signal-to-noise ratio which is important for applications. The coercivity defines the minimal magnetic field required to reverse the entire array. Preferably, the values of $H_C$ should be small so that the system of particles can be reversed with the fields that do not cause significant disturbance to the neighboring arrays. With this in mind, one can formulate the optimization criteria for the system of ferromagnetic particles as finding the optimal grid constant $d_{OPT}$ and the array size $N$ ensuring the lowest coercivity $H_C$ together with significantly high remanence $m_{x0}$. This task should be performed for the particles arranged into square, triangular and hexagonal grids.

Another important question concerns the repeatability of the hysteresis curves. To ensure this, we studied the system during eight full field cycles, constructing the histograms of the coercivity and the remanence. If the hysteresis curves were practically the same from one cycle to another, the corresponding histograms exhibit sharp peaks with the height being equal to the number of field cycles. In contrast, for hysteresis curves with poor repeatability the corresponding histograms have multiple peaks of lower intensity.

\begin{figure*}
\includegraphics[scale=0.8]{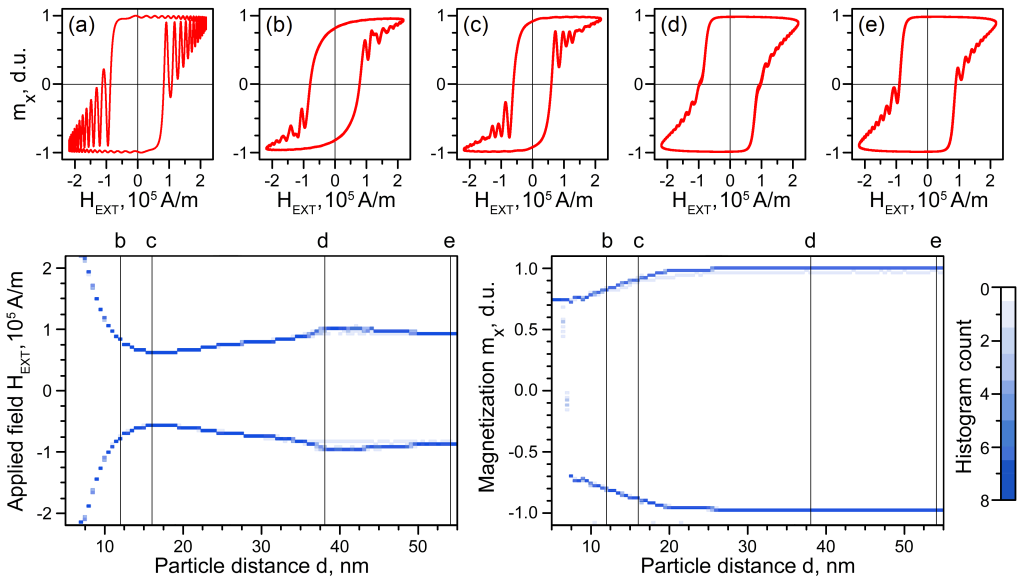}
\label{Fig2}
\caption{(Color online) Characteristic hysteresis curves for magnetization reversal in a square array of ferromagnetic particles (upper panels), with a reference hysteresis of a single particle given in the panel a). The hysteresis curves were calculated for the grid parameters: b) $d = 12$~nm, c) $d = d_{OPT} = 16$~nm, d) $d = 38$~nm and e) $d = 54$~nm. Bottom panels show the  histograms for coercivity (lower left) and remanence (lower right).}
\end{figure*}

The use of histograms is also beneficial for simplification of the optimization task. In this way, the analysis of thousands of hysteresis curves is reduced to study of histogram plots calculated by varying the inter-particle distance $d$ for the fixed $N$. Fig. 2 illustrates this approach for an array with square grid ($N=35$, $N_P = 817$). The most characteristic hysteresis curves are given in the upper part of Fig. 2 together with a reference curve obtained for a single particle (Fig. 2a). As one can see, a single macrospin does not perform well under the field frequency of 0.5 GHz. The observable magnetization precession is caused by the small value of damping coefficient $\alpha$, for which it will be desirable to use a slower field variation in order to achieve full magnetization saturation at $m_x = \pm 1$. In contrast, the particles arranged into an array respond much better to the same field frequency, clearly reaching the maximum possible remanence value (Fig. 2d,e). The coercivity for a single particle, $H_{C0} = 87$ kA/m, was used as a reference value for the further analysis.

As one can see from Fig. 2b, for $d$ = 12 nm (corresponding to the diameter of a nano-particle) the coercivity of the array coincides with that of a single particle $H_C = H_{C0}$; the remanence is $m_{x0} = 0.85$. Smaller values of $d$ will correspond to the case of overlapping particles, for which the coercivity grows abruptly. This case is beyond the scope of the present paper. However, we would like to emphasize that the obtained results are qualitatively correct, since the system of overlapping particles is expected to be more magnetically stiff in comparison to the dispersed particles because of the stronger magnetic interactions.

Increasing $d$ from 12 nm onwards one can observe a clear decrease of coercivity, which eventually reaches a minimum of $H_C = 60.2$ kA/m at the optimal distance $d_{OPT} = 16$ nm (Fig. 2c). The corresponding remanence value is considerably high ($m_{x0} = 0.90$), proving that magnetic moments of the particles $\vec{m}_i$ are almost parallel to each other. For $d > d_{OPT}$ coercivity increases again, eventually exceeding the value $H_{C0}$ as illustrated in Fig. 2d, $H_C = 100$ kA/m. The blurring of the histogram peaks observed when $d$ varies from 36 to 50 nm is due to the fact that the precessional relaxation of the magnetization starts immediately at $H_C$, so that the hysteresis curve crosses the line $m_x = 0$ several times. The pronounced coercivity jump at $d = 38$ [nm] suggests that for this value the long-range interaction field $H_{i DDI}$ ceases to be dominant in Eq. (5). The decay of the interaction magnitude is witnessed by the smooth decrease of the coercivity until it reaches the single-particle value $H_{C0}$ at $d = 54$ nm (Fig. 2e). Magnetic interactions become negligible for larger inter-particle distances, so that magnetization dynamics of each macrospin becomes independent.

\begin{figure}[!t]
\includegraphics[scale=0.8]{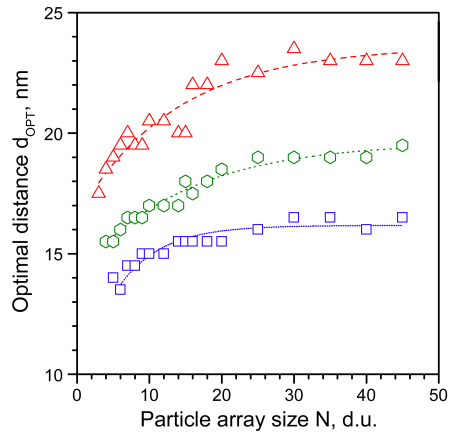}
\label{Fig3}
\caption{(Color online) Dependence of the optimal inter-particle distance $d_{OPT}$ on the array size $N$. The grid types - triangular, hexagonal and square (cf. Fig. 1) -  are denoted by the corresponding symbols. The curves are given as eye guides only.}
\end{figure}

\begin{figure}[!t]
\includegraphics[scale=0.8]{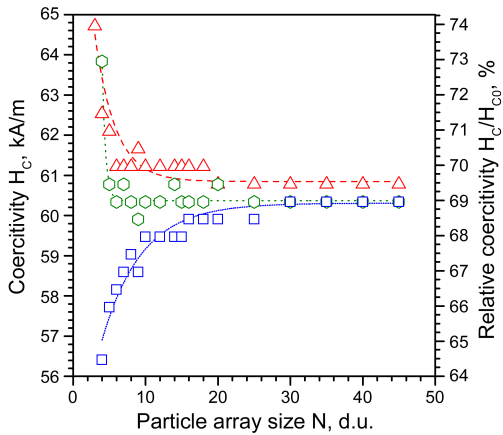}
\label{Fig4}
\caption{(Color online) Dependence of coercivity $H_C$ on array size $N$ for different grid types denoted by the corresponding symbols. The relative coercivity is given on the right-hand axis. The curves are plotted as eye guides only.}
\end{figure}

To study the influence of the array size $N$ on the properties of the system, we calculated the optimal distance $d_{OPT}$ for three grid types (Fig. 3). The parameter $N$ varied from 3 to 45, corresponding to arrays composed of 9 -- 2025 particles (cubic and  triangular grids) and 6 -- 1350 particles located in the nodes of a hexagonal grid. As one can see from the figure, the dependence of $d_{OPT}(N)$ can be generally approximated with the function $f = f_0 + A e^{-kN}$, fitting the coefficients $f_0, A$ and $k$ with the Levenberg-Marquardt method \cite{Press2007}. For considerably large systems with $N > 20\div 35$ the value of $d_{OPT}$ saturates; it is important to emphasize that for each grid type saturation value is different: $d_{OPT} = 16$ nm for particles arranged into a square grid, $d_{OPT} = 19$ nm for hexagonal grid and $d_{OPT} = 23$ nm for a triangular grid (Fig. 3). For smaller arrays ($N$ = 10) the difference between optimal grid constants is still pronounced: $d_{OPT} =15$ nm for square, $d_{OPT} = 16.5$ nm for hexagonal and $d_{OPT} = 20.5$ nm for a triangular grid, respectively.

As one can see, the triangular grid favors larger particle separation that can be explained by the larger number of the nearest neighbors (six) setting considerable restrictions on the particle dynamics. The situation is different for hexagonal and square particle arrangements. As they have a similar number of the nearest neighbors (three and four, respectively), the second- and the third-order neighbors also appear to make important contributions to the $\vec{H}_{i DDI}$. As a result, the square grid favors smaller inter-particle distances in comparison with the hexagonal one. The corresponding coercivity gain for the different array sizes is shown in Fig. 4. To simplify comparisons, the right axis gives the relative coercivity calculated with respect to the single-particle value $H_{C0}$. As one can see, all particle arrays studied featured $lower$ coercivity values compared with the single macrospin, paving promising ways towards performance improvement of spintronic  devices. Similarly to the situation observed in Fig. 3, all three curves saturate for $N > 25$, essentially setting limits for the particle array size. Indeed, despite the optimal inter-particle distance may vary with increasing of $d$ (Fig. 3), no further improvement is achieved concerning the minimization of $H_C$. The difference between the coercivity values at $N=25$ is small: $H_C =$ 60 kA/m for the square grid, $H_C =$ 60.4 kA/m for the hexagonal and $H_C =$ 60.8 kA/m for the triangular grid. However, for small arrays ($N=4$) the difference is more pronounced: 56.4 A/m for the square grid, 63.8 kA/m for the hexagonal and 64.8 kA/m for the triangular grid, which corresponds to $64\%$, $73\%$ and $74\%$ of $H_{C0}$, respectively.

\begin{figure*}[!t]
\includegraphics[scale=0.8]{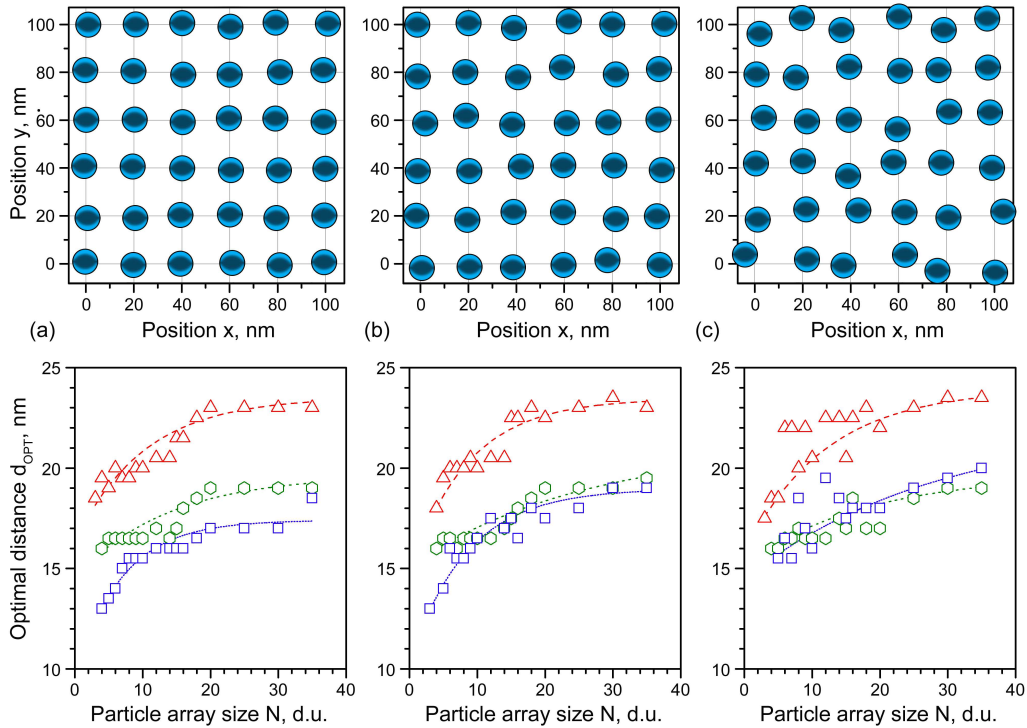}
\label{Fig5}
\caption{(Color online) The influence of particle misalignment $\Delta d$ on the optimal grid parameter $d_{OPT}(N)$ corresponding to the minimum coercivity (top view). The upper panels illustrate misalignment in square grid with a) $\Delta d = 5\%$, b) $\Delta d = 10\%$ and c) $\Delta d = 20\%$. The lower panels present $d_{OPT}(N)$ plots for the corresponding particle misalignment. The curves' symbols as in Fig. 3.}
\end{figure*}

Despite the considerable technological progress achieved in improving the precision of nano-particle placement \cite{Cao04}, it is important to know the acceptable degree of particle misalignment $\Delta d$ for which the reduction of the coercivity still occurs. To study this we performed the calculations for systems with random displacement of ferromagnetic particles from their nodes by 5\%, 10\% and 20\% of the grid parameter $d$ (Fig. 5). As one can see, the array with $5\%$ displacement (Fig. 5a, upper left panel) is quite similar to the unperturbed system. For the case of 20\% displacement (Fig. 5c, upper right panel) the array is so disordered that it becomes difficult to identify the type of the original particle arrangement. Indeed, rectangular formations give space to polygonal ones, such as irregular hexagons seen in the upper left corner of the figure (Fig. 5c). Analyzing the $d_{OPT}(N)$ scatter plots, one can see that the particles forming triangular grid keep the optimal distance at the values corresponding to the unperturbed system even under significant displacement $\Delta d$. This robustness regarding random particle displacements can be attributed to the higher magnetic stiffness. In contrast, for hexagonal and square grids the values of $d_{OPT}$ change so that for $\Delta d = 10\%$ the coercivity for both systems is almost indistinguishable.

\section{Conclusions}

The study of two-dimensional arrays of ferromagnetic nano-particles allowed the determination of the optimal grid parameter $d_{OPT}$ that is distinctly different for triangular, hexagonal and square grids. For larger inter-particle distances triangular grid is preferable; closely-packed particle arrays should rather have square grid arrangement. The reduction of the coercivity was observed for all three grid types, being more prominent for small particle arrays. For the arrays larger than 25 $\times$ 25 sites the value of $H_C$ tends to saturation. Each grid type features a distinct tolerance to random displacements of the particles. Triangular grid, being the most magnetically stiff among the systems studied, maintains the optimal inter-particle distance at $20\div 24$ nm even under considerable particle misplacements (up to $20\%$ of $d$), whereas hexagonal and square particle arrangements were less stable, degrading to $d_{OPT} \approx 16-19$ nm for $\Delta d = 10\%$. These results, to our opinion, offer several useful insights for design enhancements of ferromagnetic particle arrays aiming to achieve the best performance of spintronic devices based on them.

From our pilot calculations at different frequencies of the external magnetic field it follows that hysteresis curves lose typical oscillations next to the saturation points (Fig. 2). Thus, we expect a certain smoothing of hysteresis in the low sub-gigahertz regime.

Finally, we note that the effect of finite temperature on the overall results and in particular effects of {\it{superparamagnetic}} behavior \cite{Nowa01,KrHe09,CoKa12} deserve a separate detailed study. It is expected, however, that elevated temperatures assist the switching and the major effect is a lowering of the coercivity with increasing temperature.

\section{Acknowledgements}

The discussion with M. S{\'a}nchez-Dom{\'i}nguez on possible experimental realization of particle arrays is greatly acknowledged. This research was in part supported by CONACYT of Mexico as the Basic Science Project 129269 and by the grant from the German Research Foundation (No. SU 690/1-1).


\begin{thebibliography}{40}
\bibitem{GrSc86}
P. Gr\"unberg, R. Schreiber, Y. Pang, M.B. Brodsky, H. Sowers, Phys. Rev. Lett. {\bf 57}, 2442 (1986).
\bibitem{Baibich1988}
M.N. Baibich, J.M. Broto, A. Fert, F. Nguyen Van Dau, E. Petroff, P. Eitenne, G. Creuzet, A. Friederich, J. Chazelas, Phys. Rev. Lett. {\bf 61}, 2472 (1988).
\bibitem{Kaitsu2006}
I. Kaitsu, R. Inamura, J. Toda, T. Morita, Fujitsu Sci. Tech. J. {\bf 42}, 122 (2006).
\bibitem{Sousa2005}
R.C. Sousa, I.L. Prejbeanu, C.R. Physique {\bf 6}, 1013 (2005).
\bibitem{Kiselev2003}
S.I. Kiselev, J.C. Sankey, I.N. Krivorotov, N.C. Emley, R.J. Schoelkopf, R.A. Buhrman, D.C. Ralph, Nature {\bf 425}, 380 (2003).
\bibitem{Kaka2005}
S. Kaka, M.R. Pufall, W.R. Rippard, T.J. Silva, S.E. Russek, J.A. Katine, Nature {\bf 437}, 389 (2005).
\bibitem{Zutic2004}
I. $\check{\mathrm{Z}}$utic, J. Fabian, S. Das, Rev. Mod. Phys. {\bf 76}, 323 (2004).
\bibitem{Stiles2006}
M.D. Stiles, J. Miltat, Top. Appl. Phys. {\bf 101}, 1 (2006).
\bibitem{Silva2008}
T.J. Silva, W.H. Rippard, J. Magn. Magn. Mater. {\bf 320}, 1260 (2008).
\bibitem{Fricke2012}
L. Fricke, S. Serrano-Guisan, H.W. Schumacher, Physica B {\bf 407}, 1153 (2012).
\bibitem{Bazaliy2004}
Ya.B. Bazaliy, B.A. Jones, S.C. Zhang, Phys. Rev. B {\bf 69}, 094421 (2004).
\bibitem{Xiao2005}
J. Xiao, A. Zangwill, M.D. Stiles, Phys. Rev. B {\bf 72}, 14446 (2005).
\bibitem{Scholz2003}
W. Scholz, J. Fidler, T. Schrefl, D. Suess, R. Dittrich, H. Forster, V. Tsiantos, Comput. Mater. Sci. {\bf 28}, 366 (2003).
\bibitem{HuSc98}
A. Hubert, R. Schaefer, {\sl Magnetic Domains}, Springer, Berlin Heidelberg (1998).
\bibitem{Perry2007}
M.A. Perry, T.J. Flack, D.K. Koltsov, M.E. Welland, J. Magn. Magn. Mater. {\bf 314}, 75 (2007).
\bibitem{Dantas2011}
C.C. Dantas, Physica E {\bf 44}, 675 (2011).
\bibitem{Braun2009}
K.F. Braun, S. Sievers, M. Albrecht, U. Siegner, K. Landfester, V. Holzapfel, J. Magn. Magn. Mater. {\bf 321}, 3719 (2009).
\bibitem{Berger2012}
A. Berger, Physica B {\bf 407}, 1322  (2012).
\bibitem{Cao04}
G. Cao, {\sl Nanostructures and Nanomaterials: Synthesis, Properties and Applications}, Imperial College Press, London (2004).
\bibitem{Osbo45}
J.A. Osborn, Phys. Rev. {\bf 67}, 351 (1945).
\bibitem{Brow62}
W.F. Brown, {\sl Magnetostatic Priciples in Ferromagnetism}, North-Holland, Amsterdam (1962).
\bibitem{Coey10}
J.M.D. Coey, {\sl Magnetism and Magnetic Materials}, Cambridge University Press, New York (2010).
\bibitem{BeGr03}
M. Beleggia, M. De Graf, J. Magn. Magn. Mater. {\bf 263}, L1 (2003).
\bibitem{BeGr05}
M. Beleggia, M. De Graf, J. Magn. Magn. Mater. {\bf 285}, L1 (2005).
\bibitem{BeGr05a}
M. Beleggia, M. De Graf, Y.T. Millev, D.A. Goode, G. Rowlands, J. Phys. D: Appl. Phys. {\bf 38}, 3333 (2005).
\bibitem{BeGr06}
M. Beleggia, M. De Graf, Y.T. Millev, J. Phys. D: Appl. Phys. {\bf 39}, 891 (2006).
\bibitem{HoHe03}
J.-K. Ha, R. Hertel, J. Kirschner, Phys. Rev. B {\bf 67}, 224432 (2003).
\bibitem{ChHe05}
S. Cherifi, R. Hertel, J. Kirschner, H. Wang, R. Belkhou, A. Locatelli, S. Heun, A. Pavlovska, E. Bauer, J. Appl. Phys. {\bf 98}, 043901 (2005).
\bibitem{LaLi35}
L.D. Landau, E.M. Lifshitz, Phys. Z. Sowjetunion {\bf 8}, 153 (1935).
\bibitem{Gilb55}
T.L. Gilbert, Phys. Rev. {\bf 100}, 1243 (1955) (abstract only); IEEE Trans. Magn. {\bf 40}, 3443 (2004).
\bibitem{Press2007}
W.H. Press, S.A. Teukolsky, W.T. Vetterling, B.P. Flannery, {\sl Numerical Recipes: The Art of Scientific Computing,} Third Edition, Cambridge: Cambridge University Press (2007).
\bibitem{Nowa01}
U. Nowak, in {\it Annual Reviews of Computational Physics IX}, ed. by D. Stauffer (World Scientific, Singapore, 2001), p. 105.
\bibitem{KrHe09}
S. Krause, G. Herzog, T. Stapelfeldt, L. Berbil-Bautista, M. Bode, E.Y. Vedmedenko, and R. Wiesendager, Phys. Rev. Lett. {\bf 103}, 127202 (2009).
\bibitem{CoKa12}
For a recent review, see W.T. Coffey and Y.P. Kalmykov, J. Appl. Phys. {\bf 112}, 121301 (2012).

\end{thebibliography}
\end{document}